\documentclass[a4paper,12pt,fleqn]{scrartcl}
\title{Hempel's dilemma and the physics of computation\footnote{
published in: {\em Knowledge in Ferment: Dilemmas in Science, Scholarship and Society} (Leiden University Press, 2007).}
}
\author{Carlo Beenakker\\
Instituut-Lorentz, Universiteit Leiden}
\date{}
\begin{document}
\maketitle
\section*{The horns of the dilemma}

Carl Gustav Hempel (1905--1997) formulated the dilemma that carries his name in an attempt to determine the boundaries of physics \cite{ref1}. Where does physics go over into metaphysics? Hempel saw himself confronted with two equally unappealing choices. Is physics what physicists have discovered so far or is physics what physicists will eventually discover? The bull of Hempel's dilemma charges --- which horn will catch you?
\begin{itemize}
\item The first horn: metaphysics is what current physics cannot explain. With this choice the boundary is clear, but it is likely that some phenomena that you would classify as metaphysical will eventually find a sound physical explanation. According to 20th century quantum physics an object can be at two positions simultaneously. Should a 19th century scientist have classified this as a case of magic bilocation?
\item The second horn: metaphysics is what future physics will not be able to explain. With this choice the boundary is so vague that it is not of much use. Who knows what discoveries some 21st century Einstein will make --- so where should we draw the line today?
\end{itemize}
The purpose of this contribution is to indicate how a recently developed field of research, the physics of computation, might offer a new answer to this old question about the boundary between physics and metaphysics.

\section*{Matter is physical}

One hundred years ago the answer was clear: The boundary between physics and metaphysics is the boundary between matter and spirit. We can estimate reasonably well how much matter the universe contains, even if we are uncertain in what form the matter is present. Matter of any kind attracts other matter through the force of gravity. By measuring the velocities of remote galaxies we can calculate how much matter the intervening space contains ---even if not all that matter is visible. Some of that ``dark'' matter consists of known particles (such as the neutrino), but it is likely that it also contains as yet unknown particles.

If we restrict the field of physics to the study of matter, then we are done. Hempel's dilemma does not appear, because even if we do not know what new kinds of matter future physics might discover, the amount of matter in the universe is finite and (approximately) known. But this limitation of physics to materialism implies that whatever has no mass must be a spirit, which is not a tenable proposition. The particle of light called the photon certainly has no mass, but the laws of physics very well describe the properties of photons.

We might try to limit physics to the study of particles, with or without mass, but this restriction cannot be maintained either: As discovered by Hendrik Casimir the empty space, vacuum, has a dynamics of its own that can attract or repel objects. Quite possibly it is the repulsive force of vacuum that protects the universe from collapsing under its own weight. 

\section*{Information is physical}

To arrive at a new answer to the old question about the boundary between physics and metaphysics, I start from a statement of Rolf Landauer (1927---1999), a researcher at IBM who pioneered my own field of research, mesoscopic physics. ``Information is physical,'' was a favorite statement of Landauer \cite{ref2}, to emphasize that the processing of information by a computer is constrained by the laws of physics. Three known constraints are:
\begin{itemize}
\item Einstein's constraint: Information cannot be transferred at a speed greater than the speed of light $c$.
\item Landauer's constraint \cite{ref3}: Information cannot be erased without generating heat: the minimum heat production is $kT \ln 2$ per erased bit of information, at a temperature $T$.
\item Margolus \& Levitin's constraint \cite{ref4}: Information cannot be processed at a rate exceeding $4E/h$ operations per second, for an available energy $E$.
\end{itemize}
Together, these three constraints represent three major frame works of theoretical physics: Respectively, the theory of relativity, statistical mechanics, and quantum mechanics. Each constraint is governed by one of the fundamental constants of nature: The speed of light $c=2.98 \times 10^{8}$ m/s, Boltzmann's constant $k=1.4\times 10^{-23}$ Joule/Kelvin, and Planck's constant $h=6.6 \times 10^{-34}$ Joule/Hertz.

The most powerful computers available today operate far below these ultimate limits, but this may change at some point in the future. The field of quantum computation aims at constructing a computer that actually reaches the fundamental limits on information transfer, erasure, and processing \cite{ref5}. Once we have constructed this machine, we cannot do any better. It may take a decade, or a century, or a millennium --- but once our computer reaches the fundamental limits set by the constants of nature no amount of human ingenuity can produce a faster machine.

The fundamental limits to information processing provide an opening to the resolution of Hempel's dilemma, by restricting the capabilities of future physics based on our knowledge of current physics.

\section*{The universe is a computer}

In a reversal of Landauer's dictum, Seth Lloyd from MIT has argued \cite{ref6} that ``The universe is a computer.'' His argument proceeds as follows. Firstly the universe represents information, encoded in the state of each of its particles. The number of bits of information contained in the universe equals, by definition, the logarithm (base 2) of the number of distinct states that are available to it. Lloyd estimates that the universe as a whole has the capacity to store $10^{90}$ bits of information. A very large number, compared to the capacity of $10^{12}$ bits of a typical hard disk, but a finite number.

Secondly, the universe processes information in a systematic fashion. The ``operating system'', so to say, consists of the laws of physics. The particles evolve from one state to the other as a result of mutual interactions, in a way that is precisely dictated by the rules of quantum mechanics. We may not yet know these rules completely (in particular, in the case of the gravitational interaction), however the dogma of physics is that such rules exist and are knowable. The Margolus-Levitin constraint admits up to $10^{120}$ operations in the $10^{10}$ years since the birth of the universe \cite{ref6}. Again, this is a large number relative to the total number of operations performed by man-made computers (some $10^{30}$ in less than a century) --- but it is a finite number. 

If the universe is a computer, then it makes sense to use the constraints on information processing as constraints for physics as a whole --- not just as constraints on a subfield of physics.

\section*{Taking the bull by the horns}

Building on the insights of Landauer and Lloyd, I propose to resolve Hempel's dilemma with the definition:\\
{\em The boundary between physics and metaphysics is the boundary between what can and what cannot be computed in the age of the universe.}\\ 
Let us explore this boundary by inquiring on which side lie the three key ideas of metaphysics, as identified by Immanuel Kant: God, free will, and immortality of the soul.

{\em Is God physical or metaphysical?} What knowledge we have of God from the Abrahamic faiths suggests that His computational capacity is infinite. The computational capacity of the universe up to the present is certainly finite, as mentioned earlier, but what about the future? The cosmologist Frank Tipler \cite{ref7} has constructed a physical theory of God on the premise that the amount of information processed and stored between now and the final state of the universe is infinite. Indeed, if there is no limit to what the universe can compute in its life time, then there is no limit to the power of physics and the metaphysical is nonexistent. Most cosmologists, however, would argue (based on the expansion rate of the universe) that the computational capacity of the universe between its initial and finite state is finite, leaving God in the metaphysical domain.  

{\em Is free will physical or metaphysical?} Following Lloyd \cite{ref8} we characterize an act of free will as the outcome of a computation (= logical reasoning) by the human brain which is intrinsically unpredictable. It is not yet known how the brain operates, but even it were known, the unpredictability of its operation can remain because the laws of physics allow for unpredictability when referring to a single event. (Only averages over many events are predictable.) No metaphysical ingredients are needed.

{\em Is the immortal soul physical or metaphysical?} In order to be physical, the immortal soul should contain and process information beyond death, which is the erasure of most information in the organism. Estimates of the amount of information lost upon death are in the order of $10^{32}$ bits per human \cite{ref9}. Mankind as a whole has lost some $10^{43}$ bits of information over the course of 50,000 years. We know of no mechanism by which this amount of information could have survived by physical means, leaving the immortal soul in the metaphysical domain.

\end{document}